\documentclass[ALICE,manyauthors]{cernphprep}

\def\pt {\mbox{$p_{\rm T}$}\xspace}

\newcommand{\lambdaTheta}{\ensuremath{\lambda_{\theta}}}

\usepackage{datetime}
\usepackage{amsmath}
\usepackage{xcolor}
\usepackage[comma,square,numbers,sort&compress]{natbib}
\usepackage{hyperref}
\usepackage{lineno}
\usepackage{multirow}
\usepackage{graphicx}
\usepackage{placeins}
\usepackage[T1]{fontenc}
\usepackage{orcidlink}

\begin{document}%

\begin{titlepage}
\PHyear{2022}
\PHnumber{066}       
\PHdate{24 March}  

\title{Measurement of the J/$\psi$ polarization with respect to the event plane \\in Pb--Pb collisions at the LHC}
\ShortTitle{J/$\psi$ polarization with respect to the event plane in Pb--Pb collisions}  

\Collaboration{ALICE Collaboration\thanks{See Appendix~\ref{app:collab} for the list of collaboration members}}
\ShortAuthor{ALICE Collaboration} 


\begin{abstract}
We study the polarization of inclusive J/$\psi$ produced in Pb--Pb collisions at $\sqrt{s_{\rm NN}}=5.02$ TeV at the LHC in the dimuon channel, via the measurement of the angular distribution of its decay products. We perform the study in the rapidity region $2.5<y<4$, for three transverse momentum intervals ($2<p_{\rm T}<4$, $4<p_{\rm T}<6$, $6<p_{\rm T}<10$ GeV/$c$) and as a function of the centrality of the collision for $2<p_{\rm T}<6$ GeV/c. For the first time, the polarization is measured with respect to the event plane of the collision, by considering the angle between the positive-charge decay muon in the J/$\psi$ rest frame and the axis perpendicular to the event-plane vector in the laboratory system. A small transverse polarization is measured, with a significance reaching 3.9$\sigma$ at low $p_{\rm T}$ and for intermediate centrality values. The polarization could be connected with 
the behaviour of the quark--gluon plasma, formed in Pb--Pb collisions, as a rotating fluid with large vorticity, as well as with the existence of a strong magnetic field in the early stage of its formation.

\end{abstract}
\end{titlepage}
\setcounter{page}{2}
Quarkonia, bound states of a heavy quark--antiquark pair, have been studied for a long time because they give access to several features of the strong interaction that can be investigated with various complementary approaches (see Refs.~\cite{Brambilla:2010cs,Andronic:2015wma} for  comprehensive reviews). Calculations based on the Quantum Chromodynamics (QCD) theory  formulated on a discrete lattice~\cite{Davies:2003ik} 
can reproduce the rich spectroscopy of the various states corresponding to different radial and angular excitations of the quarkonium wave function.
The Non-Relativistic QCD (NRQCD) approach~\cite{Bodwin:1994jh} represents the most advanced tool for our understanding of quarkonium production in proton--proton collisions and is able to reproduce the measured cross sections for most states. The produced quarkonia may also exhibit polarization, defined as the alignment of the particle spin with respect to a chosen axis~\cite{Faccioli:2010kd}. The polarization can be calculated in the framework of NRQCD, and although for some states discrepancies between theory and experiment persist until today, a reasonable understanding of quarkonium production and polarization has been reached~\cite{Bodwin:2015iua,Ma:2018qvc,Feng:2018ukp}. Other approaches, such as the  Improved Color Evaporation Model (ICEM)~\cite{Ma:2016exq}, are shown to reproduce quarkonium measurements at collider energies fairly well.

Quarkonium states may also be used as a probe of the environment in which they are created or they traverse during their evolution. Their binding energy and, more generally, their spectral functions may be altered~\cite{Matsui:1986dk,Rothkopf:2019ipj} due to the presence of a quark--gluon plasma (QGP), a high energy-density state of strongly interacting matter formed in ultrarelativistic heavy-ion collisions and currently studied at RHIC and the LHC (at center-of-mass energies per nucleon--nucleon collision, $\sqrt{s_{\rm NN}}$, up to 0.2 and 5.02 TeV, respectively). These hot matter effects may lead to the dissociation or prevent the formation of the bound q${\rm \overline q}$ state. 
Furthermore, charmonia can also be significantly regenerated in the QGP phase and/or when the QGP hadronizes~\cite{BraunMunzinger:2000px,Thews:2000rj}, in particular when  the initial multiplicity of produced charm quarks is large (e.g., $>10^2$ for central Pb--Pb collisions at the LHC). Experimental results~\cite{NA50:2006yzz,CMS:2017uuv,Sirunyan:2018nsz,Acharya:2019iur} have by now  confirmed this picture.

In addition to the quarkonium yield modifications, the polarization of surviving quarkonia might be altered because of other specific features of the QGP environment. 
In particular, the fast motion of the charges of the nuclei can produce a magnetic field oriented perpendicular to the reaction plane, defined by the vector of the impact parameter of the collision and the beam direction, possibly exceeding 10$^{20}$ Gauss at LHC  energies~\cite{Skokov:2009qp,Deng:2012pc,Christakoglou:2021nhe}. The maximum value of the field  increases with energy (by a factor $\sim 10$ between RHIC and the LHC), is reached very shortly ($\ll 1$ fm/$c$) after collision time~\cite{Skokov:2009qp}, and decreases by several orders of magnitude by $t=1$ fm/$c$~\cite{Kharzeev:2007jp}. However, due to the formation of a QGP and to its finite electrical conductivity, large magnetic field values may be sustained along its entire lifetime. The production of a heavy quark pair also happens early in the collision history, within typical timescales of the order of $t\sim 1/(2m_{\rm q})\sim 0.1$ fm/$c$~\cite{Das:2016cwd}, and with the subsequent evolution toward a bound state also occurring on a time range $<1$ fm/$c$~\cite{Hufner:2000jb,Kharzeev:1999bh}, implying that polarization of charmonia may be influenced by the presence of the strong magnetic field generated in the collisions.

Another effect that may alter the polarization of quarkonia, via spin-orbit coupling, is the generation of a huge orbital angular momentum of the medium, again directed along the perpendicular to the reaction plane~\cite{Liang:2004ph,Becattini:2007sr}. In the hydrodynamic description of the QGP, this amounts to the creation of a rotating fluid with a large vorticity, with estimated values up to $\sim 10^{22}$ s$^{-1}$~\cite{STAR:2017ckg}, much larger than any other fluid existing in the universe.

Measured effects that may be related to strong e.m. fields and/or vorticity include the polarization of $\Lambda$ hyperons~\cite{STAR:2017ckg,Acharya:2019ryw}, discovered by STAR, and among vector mesons (spin quantum number equal to unity) a spin alignment of  the $\phi$ and K$^{*0}$, observed by the ALICE~\cite{Acharya:2019vpe} and STAR~\cite{STAR:2022fan} experiments. 
These hadrons are expected to be formed, up to a few GeV/$c$ transverse momentum, by light and strange quarks produced in the QGP, via recombination  processes occurring close in time to the hadronization transition. The charmonium vector mesons produced 
by regeneration effects, in particular at low $p_{\rm T}$, may therefore also exhibit spin-alignment effects as it is the case for light vector mesons. These effects can be parameterized in terms of the $\rho_{\rm 00}$ element of the spin-density matrix~\cite{Schilling:1969um}.
Because of angular momentum conservation, a net polarization of a particle sample induces an asymmetry in the angular distribution of the decay products. For the two-body dilepton decay of a vector meson, this distribution is given by 

\begin{equation}
W(\theta) \propto \frac{1}{3+\lambda_{\theta}} \left( 1 + \lambda_{\theta}\cos^2{\theta} \right) ,
\label{eq:1}   
\end{equation}

where $\theta$ is the polar emission angle of the positively charged decay lepton, with respect to a chosen axis~\cite{Faccioli:2010kd}. It can be shown that  $\lambda_{\theta}\propto (1-3\rho_{\rm 00})/(1+\rho_{\rm 00})$~\cite{Faccioli:2022polar}, so that the finite spin-alignment condition $\rho_{\rm 00}\neq 1/3$ is equivalent to the finite polarization condition $\lambda_\theta\ne 0$.

In this Letter, we report the first measurement of the J/$\psi$ polarization  with respect to an axis perpendicular to the event plane, an experimental estimator of the reaction plane, carried out by ALICE in Pb--Pb collisions at $\sqrt{s_{\rm NN}}= 5.02$ TeV. The results refer to inclusive J/$\psi$, i.e., both prompt (direct production and contribution from decays of higher-mass charmonium states) and non-prompt (from decays of hadrons containing a b quark), with the latter accounting for less than 15\% in the covered $p_{\rm T}$ range~\cite{LHCb:2011zfl}. The only previously published result on J/$\psi$ polarization for this collision system was also obtained by ALICE~\cite{Acharya:2020xko}, by measuring, via the decay J/$\psi\rightarrow\mu^+\mu^-$, the J/$\psi$ polarization in 
the helicity 
and Collins--Soper 
reference frames. These measurements showed deviations from $\lambda_\theta=0$ with a $\sim 2.1\sigma$ maximum significance at low $p_{\rm T}$, for both reference frames. In these two reference frames the polarization was measured with respect to directions directly connected with the production process, i.e., the momentum direction of the J/$\psi$ itself (helicity) or the direction of motion of the colliding hadrons (Collins-Soper). By measuring the polarization with respect to the estimated reaction plane of the nuclear collision, as done in this analysis, one rather selects a reference frame that should naturally be connected with the observation of polarization effects due to the presence of early electromagnetic fields and/or QGP vorticity. 

The data analyzed in this Letter were collected by ALICE in 2015 and 2018, and the J/$\psi$ decay to muon pairs was studied in the muon spectrometer, which covers the pseudorapidity region $-4<\eta<-2.5$. This detector consists of a 3~Tm dipole magnet, a system of five tracking (Cathode Pad Chambers) and two triggering stations (Resistive Plate Chambers), and two hadron absorbers. It is described in detail in Refs.~\cite{Aamodt:2008zz,Abelev:2014ffa}. The other detectors used for this analysis are: (i) the two layers of the Silicon Pixel Detector, SPD ($|\eta|<2$ and $|\eta|<1.4$), which represent the innermost part of the ALICE central barrel and are used for the determination of the position of the primary interaction vertex and the estimate of the event plane of the collision; (ii) the two V0 scintillator arrays ($-3.7<\eta < -1.7$ and $2.8 <\eta< 5.1$), which provide the minimum bias (MB) trigger, given by a coincidence of signals from their two sides, and are used for the rejection of beam--gas interactions. They are also used for the determination of the centrality of the collisions (see below) and for the estimate of the resolution of the event-plane determination. 

The analysed events were recorded using a dimuon trigger, defined  as the coincidence of a MB trigger together with the detection of two opposite-sign candidate tracks in the triggering system of the muon spectrometer. The trigger algorithm applies a non-sharp $p_{\rm T}$ cut, which has 50\% efficiency at 1 GeV/$c$ and becomes fully efficient ($>98$\%) beyond $p_{\rm T}\sim 2$ GeV/$c$. Selection criteria were applied at the single muon and muon pair level (see Refs.~\cite{ALICE:2018bdo,Acharya:2020xko,Acharya:2019iur} for details). 
Opposite-sign dimuons were selected in the rapidity interval  $2.5<y<4$~\footnote{Because of the symmetry of the collision system, a positive notation was adopted.} and invariant mass range $2.1<m_{\mu\mu}<4.9$ GeV/$c^2$.
The events were classified from central to peripheral according to the decreasing energy deposition in the V0 system, which is directly connected to the degree of geometric overlap of the colliding nuclei~\cite{ALICE:2013hur}. For the analysis the most central 90\% of the inelastic hadronic cross section was selected, which ensures full efficiency of the MB selection. 

The event-plane angle was estimated, event per event, as the second harmonic symmetry plane of charged particles at midrapidity,  $\Psi_{\rm 2}=\tan^{-1}(Q_{\rm 2,y}/Q_{\rm 2,x})/2$, where the transverse components of the flow vector $Q_2$ were obtained as $Q_{\rm 2,x}=\sum_{i}\cos(2\varphi_{\rm i})$ and $Q_{\rm 2,y}=\sum_{i}\sin(2\varphi_{\rm i})$, with $\varphi_{\rm i}$ being the azimuthal angle, in the center-of-mass frame of the collision, of the i$^{\rm th}$ tracklet defined by combinations of hits in the SPD. A recentering procedure~\cite{Selyuzhenkov:2007zi} was performed, as a function of the longitudinal position of the primary vertex, to remove non-uniformities in the SPD  acceptance.

Each dimuon was weighted by the inverse of the product of its acceptance times reconstruction efficiency ($A\times\epsilon$), assuming it comes from the decay of a J/$\psi$. A Monte Carlo simulation was used for the calculation of $A\times\epsilon$, with the generated J/$\psi$ signal being injected inside real MB events, to properly
reproduce the effect of detector occupancy and its variation from one centrality class to another. The $y$ and $p_{\rm T}$ input distributions for the J/$\psi$  were taken from Ref.~\cite{Acharya:2019iur}. In addition, the J/$\psi$ were generated as unpolarized, i.e., a flat distribution was assumed for the cosine of the polar angle ($\theta$) distribution of their positive decay muons with respect to the perpendicular to the event plane. A significant $p_{\rm T}$ dependence of the shape of $A\times\epsilon$ as a function of $\cos\theta$ was found, and for this reason the correction was performed on a fine 2D grid in $\cos\theta$ vs $p_{\rm T}$. Thanks to a narrow binning that leads to a small variation of these variables in each cell, the corresponding $A\times\epsilon$ values were found to be only minorly sensitive to variations in the input distributions. 
Typical values of $A\times\epsilon$ are $\sim 10$\% around $\cos\theta=0$, increase by a factor 2--2.5 when $|\cos\theta|=1$ and vary by $\sim 15$\% from peripheral to central events.

\begin{figure}[!ht]
\begin{center}
\includegraphics[scale=0.50]{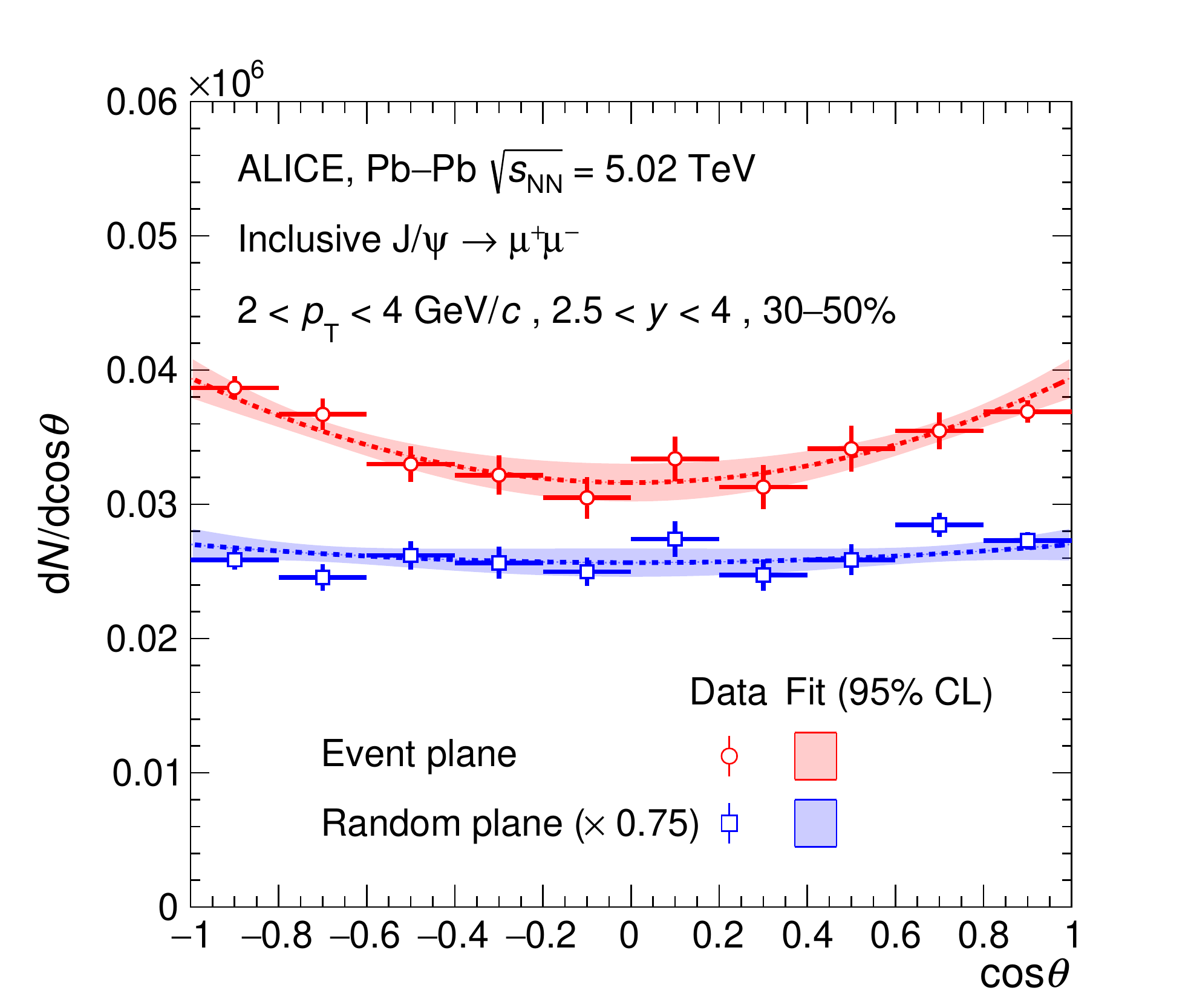}
\end{center}
\caption{Fit to the ($A\times\epsilon$)-corrected angular distribution of the positive muons from the J/$\psi$ decay, for the interval $2<p_{\rm T}<4$ GeV/$c$ and the centrality range 30--50\% (red points and curve). Only statistical uncertainties are shown for the data points. The shaded area represents the uncertainty associated with the fit. Also shown (blue points and curve) is the result of a control analysis where, for each event, the estimated event plane was rotated by a random angle.}
\label{fig:1}
\end{figure}

The extraction of the polarization parameter $\lambda_{\theta}$ was carried out as a function of centrality, for the transverse momentum interval $2<p_{\rm T}<6$ GeV/$c$, and as a function of $p_{\rm T}$ for the centrality intervals 0--20\% (most central) and 30--50\%. For each range in centrality and $p_{\rm T}$ the $A\times\epsilon$-corrected invariant mass distributions were separately obtained for ten $\cos\theta$ intervals in $-1<\cos\theta<1$. The number of J/$\psi$ for each interval was obtained by means of a $\chi^2$ minimization fit, with the signal being described by a double-sided Crystal Ball function or a pseudo-Gaussian with a mass-dependent width~\cite{ALICE-Quarkonia-signal-extraction}. The central value of the mass and the width of the J/$\psi$ were kept as free parameters of the fit, while the non-Gaussian tail parameters were fixed to the Monte Carlo values. The small contribution from the $\psi(2{\rm S})$ was included, but was found to have a negligible influence on the fit result. The background was empirically reproduced by a fourth-degree polynomial times an exponential, or a pseudo-Gaussian with a width quadratically dependent on the mass. The fits have $\chi^2$/ndf values ranging from 0.6 to 1.8. The minimum value of the signal over background ratio is 0.12 and the corresponding significance of the signal is 36, with an  increase from central to peripheral collisions and from low to high $p_{\rm T}$. Finally, the $\lambda_\theta$ values were obtained by fitting the $\cos\theta$ distributions, with $\theta$ being the angle of the positive-sign decay muon with respect to the axis perpendicular to the event plane, according to Eq.~\ref{eq:1}. In Fig.~\ref{fig:1}, an example of a fit to $A\times\epsilon$-corrected angular distributions is shown, 
together with the result of a similar analysis where for each event the event-plane angle was replaced by a randomly chosen direction. A flat angular distribution for the J/$\psi$ was obtained in the latter case. For all the studied combinations of $p_{\rm T}$ and centrality intervals, the values of $\lambdaTheta$ extracted with a random assignment of the event plane were compatible with zero, within at most 1$\sigma$. Finally, $\lambdaTheta$ must be corrected for the finite resolution on the event-plane determination. The procedure follows the one used for the K$^{*0}$ and $\phi$ mesons spin-alignment measurement~\cite{Acharya:2019vpe} which was proposed in Ref.~\cite{Tang:2018qtu}, where a simple relation between the true and observed values of the spin-density matrix element, involving the event-plane resolution, was given. The centrality-dependent resolution~\cite{ALICE:2020pvw} has a maximum value around 0.8--0.9, decreasing for very central and peripheral events, and induces a modest effect (up to +0.02) on $\lambda_\theta$.

The systematic uncertainties on $\lambda_\theta$ are related to the extraction of the J/$\psi$ signal, to the kinematic distributions used as inputs to the Monte Carlo simulation, and to the estimate of the dimuon trigger efficiency. The first source was evaluated by comparing the $\lambda_\theta$ values obtained from angular distributions extracted with different choices for the signal and background shapes in the invariant mass fits, and by using various fit ranges, from $2.1<m_{\mu\mu}<4.9$ GeV/$c^2$ (wider) to $2.5<m_{\mu\mu}<4.5$ GeV/$c^2$ (narrower). The absolute values of this systematic uncertainty, taken as the RMS of the $\lambda_\theta$ values, range between 0.02 and 0.04 as a function of centrality and from 0.02 to 0.06 as a function of $p_{\rm T}$.  
Concerning the Monte Carlo generation, 
due to suppression and regeneration effects on the J/$\psi$ yields occurring in Pb--Pb collisions~\cite{Acharya:2019iur}, the $p_{\rm T}$ and $y$ distributions have a centrality dependence. A weight to the default centrality-integrated distributions was applied in order to reproduce such dependence in the $A\times\epsilon$ calculations. The effect on the evaluation of $\lambda_{\theta}$ was found to be small, being less than 0.01 as a function of centrality, and smaller than 0.02 as a function of $p_{\rm T}$.
Since the muon trigger response function exhibits a slight difference in data and in the Monte Carlo for $\pt<2$ GeV/$\it{c}$, the $\lambdaTheta$ parameter was extracted after weighting the $A\times\varepsilon$ in order to take into account this discrepancy. The variation of the results after this correction, $\sim0.01$ as a function of centrality and 0.01--0.02 as a function of $p_{\rm T}$, was taken as the systematic uncertainty on the trigger efficiency.  
Further efficiency-related uncertainties (tracking, matching between tracks in the tracking detectors and tracklets in the trigger detectors) were found to have a negligible influence on the polarization parameters. 
The total systematic uncertainty on $\lambdaTheta$ was obtained as the quadratic sum of the values corresponding to each source, see Table~\ref{table:jpsi_systematic_uncertainties_summary}.

\begin{table}[!ht]
	\begin{center}
		\caption{Systematic uncertainties on the evaluation of the $\lambdaTheta$ parameter. The quoted uncertainties for the various sources are considered as  uncorrelated.}\label{table:jpsi_systematic_uncertainties_summary}
		\begin{tabular}{cc|ccc|c}
			\hline
			\hline
			$\pt$ (GeV/\textit{c}) & Centrality & Signal & Trigger & Input MC & Total\\
			\hline
			\multirow{4}{*}{2--6}
			& 0--20\% & 0.045 & 0.006 & 0.006 & 0.046\\ 
			& 20--40\% & 0.027 & 0.010 & 0.006 & 0.030\\ 
			& 40--60\% & 0.015 & 0.006 & 0.002 & 0.017\\ 
			& 60--90\% & 0.016 & 0.007 & 0.003 & 0.018\\
			\hline
			Centrality & $\pt$ (GeV/\textit{c}) & Signal & Trigger & Input MC & Total\\
			\hline
			\multirow{3}{*}{0--20\%}
			& 2--4 & 0.063 & 0.017 & 0.007 & 0.065\\ 
			& 4--6 & 0.020 & 0.011 & 0.007 & 0.024\\ 
			& 6--10 & 0.024 & 0.006 & 0.008 & 0.026\\
			\hline
			\multirow{3}{*}{30--50\%}
			& 2--4 & 0.032 & 0.007 & 0.006 & 0.033\\ 
			& 4--6 & 0.026 & 0.015 & 0.008 & 0.031\\ 
			& 6--10 & 0.025 & 0.006 & 0.012 & 0.029\\
			\hline
			\hline
		\end{tabular}
	\end{center}
\end{table}

\begin{figure}[!h] 
\begin{center}
\includegraphics[scale=0.39]{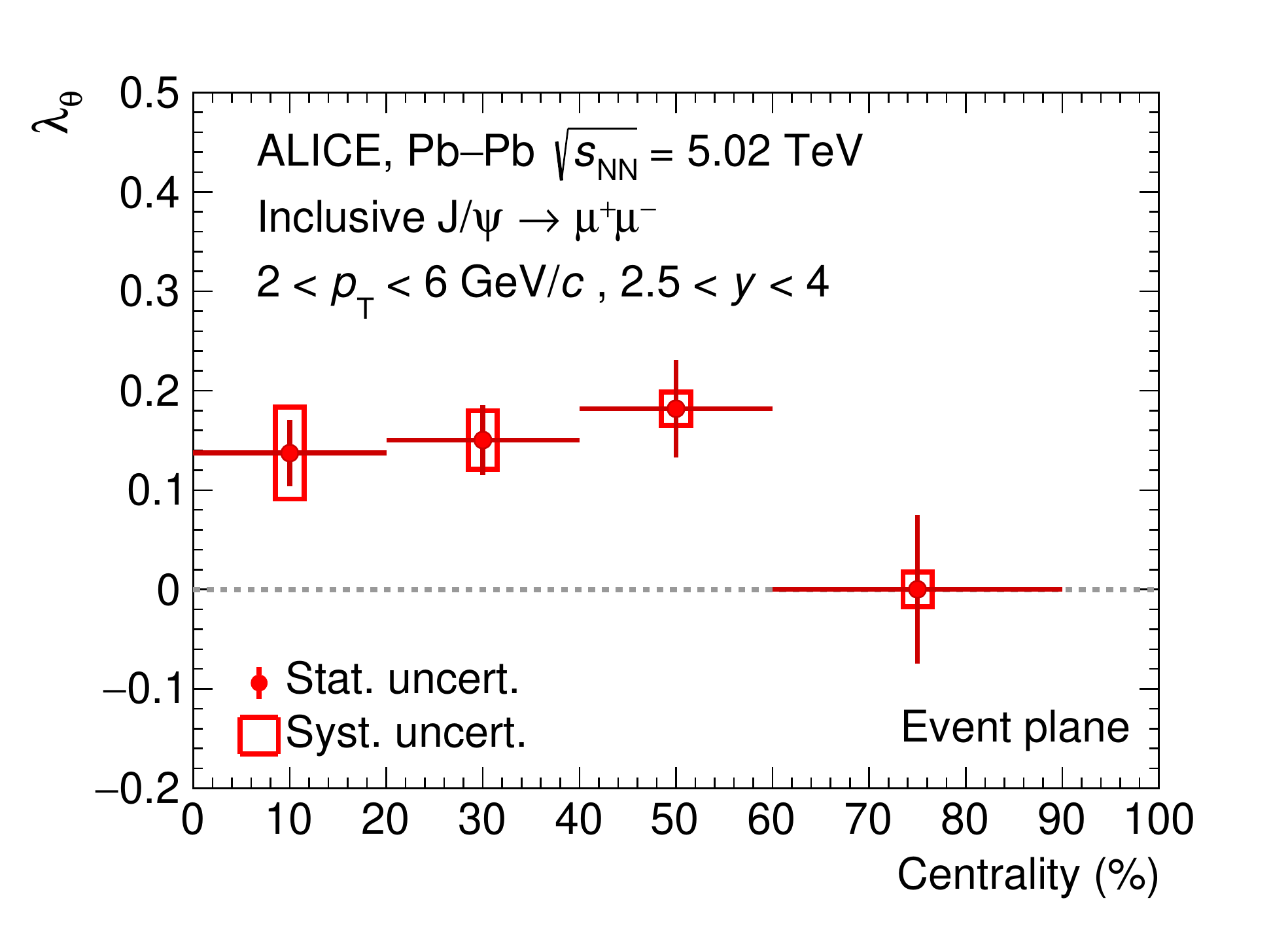}
\includegraphics[scale=0.39]{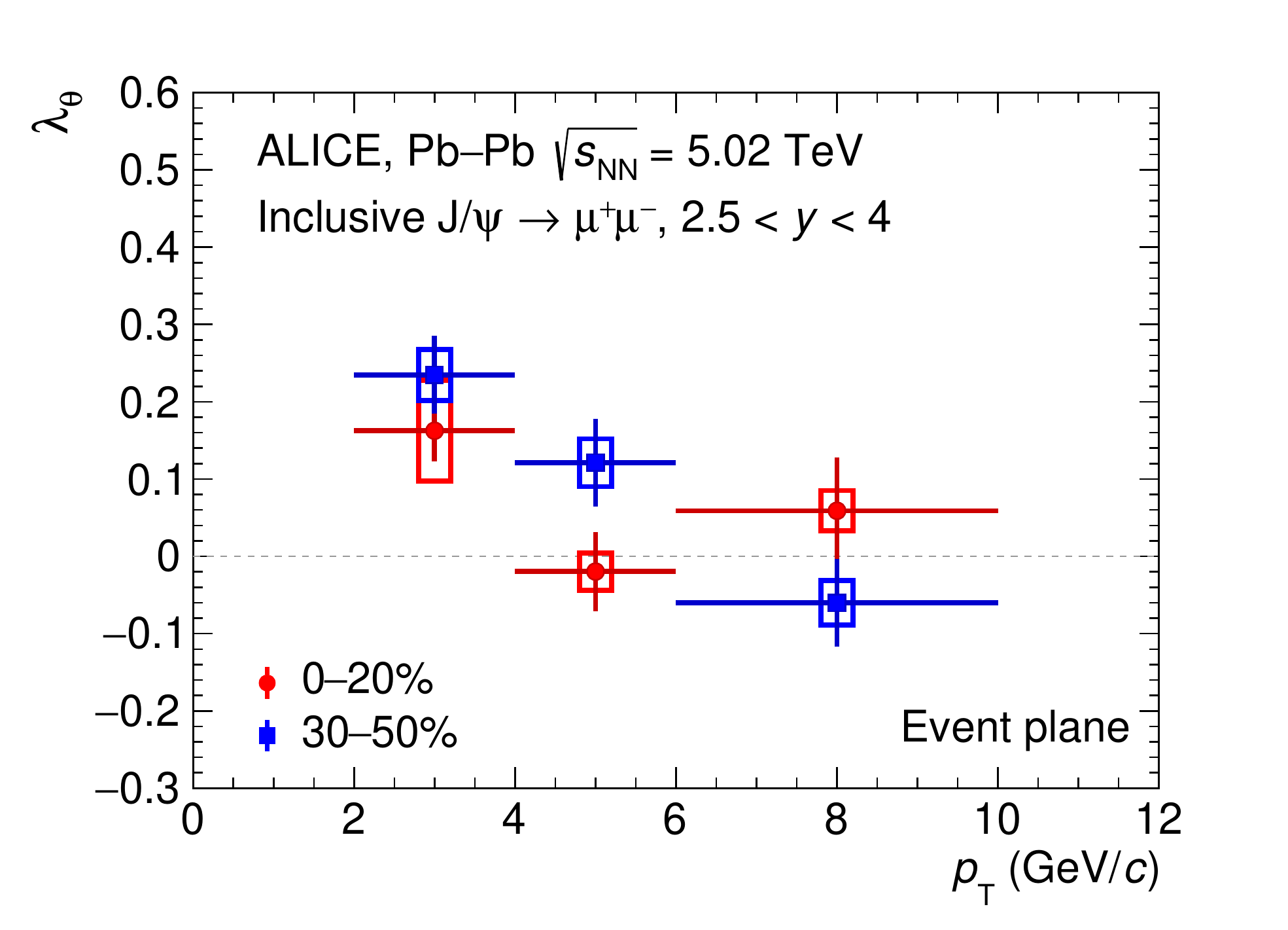}
\end{center}
\caption{Centrality (left panel) and $p_{\rm T}$ dependence (right panel) of $\lambdaTheta$. The vertical bars represent the statistical uncertainties, while the boxes correspond to the systematic uncertainties. The horizontal bars show the size of the corresponding centrality and $p_{\rm T}$ ranges, with the data points being located at the center of each interval. }
\label{fig:2}
\end{figure}

In Fig.~\ref{fig:2}, the centrality dependence of $\lambdaTheta$  for the range $2<p_{\rm T}<6$ GeV/$c$ is presented (left panel), as well as the $p_{\rm T}$ dependence  of $\lambdaTheta$ for central (0--20\%) and intermediate centrality (30--50\%) events (right panel). As a function of centrality a small but significant transverse polarization is found from central collisions down to the 40--60\% centrality interval, where a 3.5$\sigma$ effect is observed. The results as a function of $p_{\rm T}$ may indicate that the deviation from zero is larger at small transverse momentum. The maximum deviation from $\lambda_{\theta}=0$ as a function of $p_{\rm T}$ is observed for $2<p_{\rm T}<4$ GeV/$c$ and 30--50\% centrality where, considering the total uncertainty, a 3.9$\sigma$ effect is present. The results correspond to inclusive J/$\psi$ production, implying that a small contribution from a potential polarization of parent beauty hadrons, which could anyway be diluted in the decay process~\cite{Abulencia:2007us}, might be present.

Previous measurements carried out by ALICE on K$^{*0}$ and $\phi$ spin alignment~\cite{Acharya:2019vpe} had established evidence of a significant effect for vector mesons in heavy-ion collisions, stronger at low $p_{\rm T}$ and for semi-central Pb--Pb collisions. The maximum $\lambdaTheta$ value measured for the J/$\psi$ ($\sim0.2$) in this analysis would translate, in the language of spin matrix elements, to $\rho_{\rm 00}\sim 0.25$. This result implies a deviation of --0.08 from $\rho_{\rm 00}=1/3$ (corresponding to no spin alignment), in the same direction with respect to the corresponding deviations of about --0.2 for K$^{*0}$ and --0.1 for $\phi$. It can be noted that the $p_{\rm T}$ and centrality dependence of the observed spin-alignment effects are qualitatively consistent between light vector mesons~\cite{Acharya:2019vpe} and charmonia. In particular, for the centrality dependence, the possible increase  of $\lambda_\theta$ from central to semi-central collisions, followed by a decrease in peripheral events, is in qualitative agreement with the dependence of the initial angular momentum on impact parameter~\cite{Becattini:2007sr}.
The results for K$^{*0}$ and $\phi$ are consistent with a scenario of quark polarization in the presence of a large angular momentum of the system~\cite{Acharya:2019vpe}. The results shown in this Letter may confirm this interpretation also for the charmonium sector. On the other hand, charm quarks are produced early in the collision history and could be more sensitive to additional effects related to strong electromagnetic fields. Those effects would lead to a net increase of $\rho_{\rm 00}$ with respect to 1/3~\cite{Yang:2017sdk}. Our data, being roughly compatible with the result for K$^{*0}$ and $\phi$, do not give evidence for a scenario that includes a significant additional contribution to $\rho_{\rm 00}$.  
Clearly, these hints need to be confirmed by theory studies devoted to charm and charmonium production, which are currently under development~\cite{Sheng:2022wsy}. On the experimental side, significant detector upgrades and a factor $\sim 20$ increase in the available integrated luminosity in the LHC runs 3 and 4~\cite{Antonioli:2013ppp,CERN-LHCC-2015-001} will allow a decisive improvement in the statistical significance of these results as well as an extension toward midrapidity. Furthermore, a first measurement for $\Upsilon$ states, which are produced even earlier in the collision history and experience little regeneration in the QGP, could be carried out. This measurement will  potentially be more sensitive to the early strong electromagnetic fields. 

In summary, we have reported on the first measurement of the polarization for inclusive J/$\psi$ produced in Pb--Pb interactions at $\sqrt{s_{\rm NN}}=5.02$ TeV, carried out by ALICE using the direction perpendicular to the event plane of the collision as the polarization axis. This choice makes this measurement potentially sensitive to the strong magnetic field created in high-energy nuclear collisions, as well as to vorticity effects in the QGP. A small but significant transverse polarization signal, reaching 3.9$\sigma$ for $2<p_{\rm T}<4$ GeV/$c$ and 30--50\% centrality, is measured. The effect is roughly compatible with that seen for light vector mesons and does not show a significant additional contribution that may be related to the presence of a strong electromagnetic field. However, the differences in the production timescale of the involved quarks require dedicated theory studies for a quantitative understanding of this observation and a precise connection with the QGP properties at its origin.



\newenvironment{acknowledgement}{\relax}{\relax}
\begin{acknowledgement}
\section*{Acknowledgements}

The ALICE Collaboration would like to thank all its engineers and technicians for their invaluable contributions to the construction of the experiment and the CERN accelerator teams for the outstanding performance of the LHC complex.
The ALICE Collaboration gratefully acknowledges the resources and support provided by all Grid centres and the Worldwide LHC Computing Grid (WLCG) collaboration.
The ALICE Collaboration acknowledges the following funding agencies for their support in building and running the ALICE detector:
A. I. Alikhanyan National Science Laboratory (Yerevan Physics Institute) Foundation (ANSL), State Committee of Science and World Federation of Scientists (WFS), Armenia;
Austrian Academy of Sciences, Austrian Science Fund (FWF): [M 2467-N36] and Nationalstiftung f\"{u}r Forschung, Technologie und Entwicklung, Austria;
Ministry of Communications and High Technologies, National Nuclear Research Center, Azerbaijan;
Conselho Nacional de Desenvolvimento Cient\'{\i}fico e Tecnol\'{o}gico (CNPq), Financiadora de Estudos e Projetos (Finep), Funda\c{c}\~{a}o de Amparo \`{a} Pesquisa do Estado de S\~{a}o Paulo (FAPESP) and Universidade Federal do Rio Grande do Sul (UFRGS), Brazil;
Bulgarian Ministry of Education and Science, within the National Roadmap for Research Infrastructures 2020-2027 (object CERN), Bulgaria;
Ministry of Education of China (MOEC) , Ministry of Science \& Technology of China (MSTC) and National Natural Science Foundation of China (NSFC), China;
Ministry of Science and Education and Croatian Science Foundation, Croatia;
Centro de Aplicaciones Tecnol\'{o}gicas y Desarrollo Nuclear (CEADEN), Cubaenerg\'{\i}a, Cuba;
Ministry of Education, Youth and Sports of the Czech Republic, Czech Republic;
The Danish Council for Independent Research | Natural Sciences, the VILLUM FONDEN and Danish National Research Foundation (DNRF), Denmark;
Helsinki Institute of Physics (HIP), Finland;
Commissariat \`{a} l'Energie Atomique (CEA) and Institut National de Physique Nucl\'{e}aire et de Physique des Particules (IN2P3) and Centre National de la Recherche Scientifique (CNRS), France;
Bundesministerium f\"{u}r Bildung und Forschung (BMBF) and GSI Helmholtzzentrum f\"{u}r Schwerionenforschung GmbH, Germany;
General Secretariat for Research and Technology, Ministry of Education, Research and Religions, Greece;
National Research, Development and Innovation Office, Hungary;
Department of Atomic Energy Government of India (DAE), Department of Science and Technology, Government of India (DST), University Grants Commission, Government of India (UGC) and Council of Scientific and Industrial Research (CSIR), India;
National Research and Innovation Agency - BRIN, Indonesia;
Istituto Nazionale di Fisica Nucleare (INFN), Italy;
Japanese Ministry of Education, Culture, Sports, Science and Technology (MEXT) and Japan Society for the Promotion of Science (JSPS) KAKENHI, Japan;
Consejo Nacional de Ciencia (CONACYT) y Tecnolog\'{i}a, through Fondo de Cooperaci\'{o}n Internacional en Ciencia y Tecnolog\'{i}a (FONCICYT) and Direcci\'{o}n General de Asuntos del Personal Academico (DGAPA), Mexico;
Nederlandse Organisatie voor Wetenschappelijk Onderzoek (NWO), Netherlands;
The Research Council of Norway, Norway;
Commission on Science and Technology for Sustainable Development in the South (COMSATS), Pakistan;
Pontificia Universidad Cat\'{o}lica del Per\'{u}, Peru;
Ministry of Education and Science, National Science Centre and WUT ID-UB, Poland;
Korea Institute of Science and Technology Information and National Research Foundation of Korea (NRF), Republic of Korea;
Ministry of Education and Scientific Research, Institute of Atomic Physics, Ministry of Research and Innovation and Institute of Atomic Physics and University Politehnica of Bucharest, Romania;
Ministry of Education, Science, Research and Sport of the Slovak Republic, Slovakia;
National Research Foundation of South Africa, South Africa;
Swedish Research Council (VR) and Knut \& Alice Wallenberg Foundation (KAW), Sweden;
European Organization for Nuclear Research, Switzerland;
Suranaree University of Technology (SUT), National Science and Technology Development Agency (NSTDA), Thailand Science Research and Innovation (TSRI) and National Science, Research and Innovation Fund (NSRF), Thailand;
Turkish Energy, Nuclear and Mineral Research Agency (TENMAK), Turkey;
National Academy of  Sciences of Ukraine, Ukraine;
Science and Technology Facilities Council (STFC), United Kingdom;
National Science Foundation of the United States of America (NSF) and United States Department of Energy, Office of Nuclear Physics (DOE NP), United States of America.
In addition, individual groups or members have received support from:
Marie Sk\l{}odowska Curie, Strong 2020 - Horizon 2020, European Research Council (grant nos. 824093, 896850, 950692), European Union;
Academy of Finland (Center of Excellence in Quark Matter) (grant nos. 346327, 346328), Finland;
Programa de Apoyos para la Superaci\'{o}n del Personal Acad\'{e}mico, UNAM, Mexico.
\end{acknowledgement}

\bibliographystyle{utphys}  
\bibliography{bibliography}

\newpage
\appendix
\section{The ALICE Collaboration}
\label{app:collab}
\begin{flushleft} 
\small

S.~Acharya\,\orcidlink{0000-0002-9213-5329}\,$^{\rm 124,131}$, 
D.~Adamov\'{a}\,\orcidlink{0000-0002-0504-7428}\,$^{\rm 86}$, 
A.~Adler$^{\rm 69}$, 
G.~Aglieri Rinella\,\orcidlink{0000-0002-9611-3696}\,$^{\rm 32}$, 
M.~Agnello\,\orcidlink{0000-0002-0760-5075}\,$^{\rm 29}$, 
N.~Agrawal\,\orcidlink{0000-0003-0348-9836}\,$^{\rm 50}$, 
Z.~Ahammed\,\orcidlink{0000-0001-5241-7412}\,$^{\rm 131}$, 
S.~Ahmad\,\orcidlink{0000-0003-0497-5705}\,$^{\rm 15}$, 
S.U.~Ahn\,\orcidlink{0000-0001-8847-489X}\,$^{\rm 70}$, 
I.~Ahuja\,\orcidlink{0000-0002-4417-1392}\,$^{\rm 37}$
% ...

\section*{Affiliation Notes}

$^{\rm I}$ Deceased\\
$^{\rm II}$ Also at: Max-Planck-Institut f\"{u}r Physik, Munich, Germany\\
$^{\rm III}$ Also at: Italian National Agency for New Technologies, Energy and Sustainable Economic Development (ENEA), Bologna, Italy\\
$^{\rm IV}$ Also at: Dipartimento DET del Politecnico di Torino, Turin, Italy\\
$^{\rm V}$ Also at: Department of Applied Physics, Aligarh Muslim University, Aligarh, India\\
$^{\rm VI}$ Also at: Institute of Theoretical Physics, University of Wroclaw, Poland\\
$^{\rm VII}$ Also at: An institution covered by a cooperation agreement with CERN\\

\section*{Collaboration Institutes}

$^{1}$ A.I. Alikhanyan National Science Laboratory (Yerevan Physics Institute) Foundation, Yerevan, Armenia\\
$^{2}$ AGH University of Science and Technology, Cracow, Poland\\
$^{3}$ Bogolyubov Institute for Theoretical Physics, National Academy of Sciences of Ukraine, Kiev, Ukraine\\
$^{4}$ Bose Institute, Department of Physics  and Centre for Astroparticle Physics and Space Science (CAPSS), Kolkata, India\\
$^{5}$ California Polytechnic State University, San Luis Obispo, California, United States\\
$^{6}$ Central China Normal University, Wuhan, China\\
$^{7}$ Centro de Aplicaciones Tecnol\'{o}gicas y Desarrollo Nuclear (CEADEN), Havana, Cuba\\
$^{8}$ Centro de Investigaci\'{o}n y de Estudios Avanzados (CINVESTAV), Mexico City and M\'{e}rida, Mexico\\
$^{9}$ Chicago State University, Chicago, Illinois, United States\\
$^{10}$ China Institute of Atomic Energy, Beijing, China\\
$^{11}$ Chungbuk National University, Cheongju, Republic of Korea\\
$^{12}$ Comenius University Bratislava, Faculty of Mathematics, Physics and Informatics, Bratislava, Slovak Republic\\
$^{13}$ COMSATS University Islamabad, Islamabad, Pakistan\\
$^{14}$ Creighton University, Omaha, Nebraska, United States\\
$^{15}$ Department of Physics, Aligarh Muslim University, Aligarh, India\\
$^{16}$ Department of Physics, Pusan National University, Pusan, Republic of Korea\\
$^{17}$ Department of Physics, Sejong University, Seoul, Republic of Korea\\
$^{18}$ Department of Physics, University of California, Berkeley, California, United States\\
$^{19}$ Department of Physics, University of Oslo, Oslo, Norway\\
$^{20}$ Department of Physics and Technology, University of Bergen, Bergen, Norway\\
$^{21}$ Dipartimento di Fisica, Universit\`{a} di Pavia, Pavia, Italy\\
$^{22}$ Dipartimento di Fisica dell'Universit\`{a} and Sezione INFN, Cagliari, Italy\\
$^{23}$ Dipartimento di Fisica dell'Universit\`{a} and Sezione INFN, Trieste, Italy\\
$^{24}$ Dipartimento di Fisica dell'Universit\`{a} and Sezione INFN, Turin, Italy\\
$^{25}$ Dipartimento di Fisica e Astronomia dell'Universit\`{a} and Sezione INFN, Bologna, Italy\\
$^{26}$ Dipartimento di Fisica e Astronomia dell'Universit\`{a} and Sezione INFN, Catania, Italy\\
$^{27}$ Dipartimento di Fisica e Astronomia dell'Universit\`{a} and Sezione INFN, Padova, Italy\\
$^{28}$ Dipartimento di Fisica `E.R.~Caianiello' dell'Universit\`{a} and Gruppo Collegato INFN, Salerno, Italy\\
$^{29}$ Dipartimento DISAT del Politecnico and Sezione INFN, Turin, Italy\\
$^{30}$ Dipartimento di Scienze MIFT, Universit\`{a} di Messina, Messina, Italy\\
$^{31}$ Dipartimento Interateneo di Fisica `M.~Merlin' and Sezione INFN, Bari, Italy\\
$^{32}$ European Organization for Nuclear Research (CERN), Geneva, Switzerland\\
$^{33}$ Faculty of Electrical Engineering, Mechanical Engineering and Naval Architecture, University of Split, Split, Croatia\\
$^{34}$ Faculty of Engineering and Science, Western Norway University of Applied Sciences, Bergen, Norway\\
$^{35}$ Faculty of Nuclear Sciences and Physical Engineering, Czech Technical University in Prague, Prague, Czech Republic\\
$^{36}$ Faculty of Physics, Sofia University, Sofia, Bulgaria\\
$^{37}$ Faculty of Science, P.J.~\v{S}af\'{a}rik University, Ko\v{s}ice, Slovak Republic\\
$^{38}$ Frankfurt Institute for Advanced Studies, Johann Wolfgang Goethe-Universit\"{a}t Frankfurt, Frankfurt, Germany\\
$^{39}$ Fudan University, Shanghai, China\\
$^{40}$ Gangneung-Wonju National University, Gangneung, Republic of Korea\\
$^{41}$ Gauhati University, Department of Physics, Guwahati, India\\
$^{42}$ Helmholtz-Institut f\"{u}r Strahlen- und Kernphysik, Rheinische Friedrich-Wilhelms-Universit\"{a}t Bonn, Bonn, Germany\\
$^{43}$ Helsinki Institute of Physics (HIP), Helsinki, Finland\\
$^{44}$ High Energy Physics Group,  Universidad Aut\'{o}noma de Puebla, Puebla, Mexico\\
$^{45}$ Horia Hulubei National Institute of Physics and Nuclear Engineering, Bucharest, Romania\\
$^{46}$ Indian Institute of Technology Bombay (IIT), Mumbai, India\\
$^{47}$ Indian Institute of Technology Indore, Indore, India\\
$^{48}$ INFN, Laboratori Nazionali di Frascati, Frascati, Italy\\
$^{49}$ INFN, Sezione di Bari, Bari, Italy\\
$^{50}$ INFN, Sezione di Bologna, Bologna, Italy\\
$^{51}$ INFN, Sezione di Cagliari, Cagliari, Italy\\
$^{52}$ INFN, Sezione di Catania, Catania, Italy\\
$^{53}$ INFN, Sezione di Padova, Padova, Italy\\
$^{54}$ INFN, Sezione di Pavia, Pavia, Italy\\
$^{55}$ INFN, Sezione di Torino, Turin, Italy\\
$^{56}$ INFN, Sezione di Trieste, Trieste, Italy\\
$^{57}$ Inha University, Incheon, Republic of Korea\\
$^{58}$ Institute for Gravitational and Subatomic Physics (GRASP), Utrecht University/Nikhef, Utrecht, Netherlands\\
$^{59}$ Institute of Experimental Physics, Slovak Academy of Sciences, Ko\v{s}ice, Slovak Republic\\
$^{60}$ Institute of Physics, Homi Bhabha National Institute, Bhubaneswar, India\\
$^{61}$ Institute of Physics of the Czech Academy of Sciences, Prague, Czech Republic\\
$^{62}$ Institute of Space Science (ISS), Bucharest, Romania\\
$^{63}$ Institut f\"{u}r Kernphysik, Johann Wolfgang Goethe-Universit\"{a}t Frankfurt, Frankfurt, Germany\\
$^{64}$ Instituto de Ciencias Nucleares, Universidad Nacional Aut\'{o}noma de M\'{e}xico, Mexico City, Mexico\\
$^{65}$ Instituto de F\'{i}sica, Universidade Federal do Rio Grande do Sul (UFRGS), Porto Alegre, Brazil\\
$^{66}$ Instituto de F\'{\i}sica, Universidad Nacional Aut\'{o}noma de M\'{e}xico, Mexico City, Mexico\\
$^{67}$ iThemba LABS, National Research Foundation, Somerset West, South Africa\\
$^{68}$ Jeonbuk National University, Jeonju, Republic of Korea\\
$^{69}$ Johann-Wolfgang-Goethe Universit\"{a}t Frankfurt Institut f\"{u}r Informatik, Fachbereich Informatik und Mathematik, Frankfurt, Germany\\
$^{70}$ Korea Institute of Science and Technology Information, Daejeon, Republic of Korea\\
$^{71}$ KTO Karatay University, Konya, Turkey\\
$^{72}$ Laboratoire de Physique des 2 Infinis, Ir\`{e}ne Joliot-Curie, Orsay, France\\
$^{73}$ Laboratoire de Physique Subatomique et de Cosmologie, Universit\'{e} Grenoble-Alpes, CNRS-IN2P3, Grenoble, France\\
$^{74}$ Lawrence Berkeley National Laboratory, Berkeley, California, United States\\
$^{75}$ Lund University Department of Physics, Division of Particle Physics, Lund, Sweden\\
$^{76}$ Nagasaki Institute of Applied Science, Nagasaki, Japan\\
$^{77}$ Nara Women{'}s University (NWU), Nara, Japan\\
$^{78}$ National and Kapodistrian University of Athens, School of Science, Department of Physics , Athens, Greece\\
$^{79}$ National Centre for Nuclear Research, Warsaw, Poland\\
$^{80}$ National Institute of Science Education and Research, Homi Bhabha National Institute, Jatni, India\\
$^{81}$ National Nuclear Research Center, Baku, Azerbaijan\\
$^{82}$ National Research and Innovation Agency - BRIN, Jakarta, Indonesia\\
$^{83}$ Niels Bohr Institute, University of Copenhagen, Copenhagen, Denmark\\
$^{84}$ Nikhef, National institute for subatomic physics, Amsterdam, Netherlands\\
$^{85}$ Nuclear Physics Group, STFC Daresbury Laboratory, Daresbury, United Kingdom\\
$^{86}$ Nuclear Physics Institute of the Czech Academy of Sciences, Husinec-\v{R}e\v{z}, Czech Republic\\
$^{87}$ Oak Ridge National Laboratory, Oak Ridge, Tennessee, United States\\
$^{88}$ Ohio State University, Columbus, Ohio, United States\\
$^{89}$ Physics department, Faculty of science, University of Zagreb, Zagreb, Croatia\\
$^{90}$ Physics Department, Panjab University, Chandigarh, India\\
$^{91}$ Physics Department, University of Jammu, Jammu, India\\
$^{92}$ Physics Department, University of Rajasthan, Jaipur, India\\
$^{93}$ Physics Program and International Institute for Sustainability with Knotted Chiral Meta Matter (SKCM2), Hiroshima University, Hiroshima, Japan\\
$^{94}$ Physikalisches Institut, Eberhard-Karls-Universit\"{a}t T\"{u}bingen, T\"{u}bingen, Germany\\
$^{95}$ Physikalisches Institut, Ruprecht-Karls-Universit\"{a}t Heidelberg, Heidelberg, Germany\\
$^{96}$ Physik Department, Technische Universit\"{a}t M\"{u}nchen, Munich, Germany\\
$^{97}$ Politecnico di Bari and Sezione INFN, Bari, Italy\\
$^{98}$ Research Division and ExtreMe Matter Institute EMMI, GSI Helmholtzzentrum f\"ur Schwerionenforschung GmbH, Darmstadt, Germany\\
$^{99}$ Saha Institute of Nuclear Physics, Homi Bhabha National Institute, Kolkata, India\\
$^{100}$ School of Physics and Astronomy, University of Birmingham, Birmingham, United Kingdom\\
$^{101}$ Secci\'{o}n F\'{\i}sica, Departamento de Ciencias, Pontificia Universidad Cat\'{o}lica del Per\'{u}, Lima, Peru\\
$^{102}$ Stefan Meyer Institut f\"{u}r Subatomare Physik (SMI), Vienna, Austria\\
$^{103}$ SUBATECH, IMT Atlantique, Nantes Universit\'{e}, CNRS-IN2P3, Nantes, France\\
$^{104}$ Suranaree University of Technology, Nakhon Ratchasima, Thailand\\
$^{105}$ Technical University of Ko\v{s}ice, Ko\v{s}ice, Slovak Republic\\
$^{106}$ The Henryk Niewodniczanski Institute of Nuclear Physics, Polish Academy of Sciences, Cracow, Poland\\
$^{107}$ The University of Texas at Austin, Austin, Texas, United States\\
$^{108}$ Universidad Aut\'{o}noma de Sinaloa, Culiac\'{a}n, Mexico\\
$^{109}$ Universidade de S\~{a}o Paulo (USP), S\~{a}o Paulo, Brazil\\
$^{110}$ Universidade Estadual de Campinas (UNICAMP), Campinas, Brazil\\
$^{111}$ Universidade Federal do ABC, Santo Andre, Brazil\\
$^{112}$ University of Cape Town, Cape Town, South Africa\\
$^{113}$ University of Houston, Houston, Texas, United States\\
$^{114}$ University of Jyv\"{a}skyl\"{a}, Jyv\"{a}skyl\"{a}, Finland\\
$^{115}$ University of Kansas, Lawrence, Kansas, United States\\
$^{116}$ University of Liverpool, Liverpool, United Kingdom\\
$^{117}$ University of Science and Technology of China, Hefei, China\\
$^{118}$ University of South-Eastern Norway, Kongsberg, Norway\\
$^{119}$ University of Tennessee, Knoxville, Tennessee, United States\\
$^{120}$ University of the Witwatersrand, Johannesburg, South Africa\\
$^{121}$ University of Tokyo, Tokyo, Japan\\
$^{122}$ University of Tsukuba, Tsukuba, Japan\\
$^{123}$ University Politehnica of Bucharest, Bucharest, Romania\\
$^{124}$ Universit\'{e} Clermont Auvergne, CNRS/IN2P3, LPC, Clermont-Ferrand, France\\
$^{125}$ Universit\'{e} de Lyon, CNRS/IN2P3, Institut de Physique des 2 Infinis de Lyon, Lyon, France\\
$^{126}$ Universit\'{e} de Strasbourg, CNRS, IPHC UMR 7178, F-67000 Strasbourg, France, Strasbourg, France\\
$^{127}$ Universit\'{e} Paris-Saclay Centre d'Etudes de Saclay (CEA), IRFU, D\'{e}partment de Physique Nucl\'{e}aire (DPhN), Saclay, France\\
$^{128}$ Universit\`{a} degli Studi di Foggia, Foggia, Italy\\
$^{129}$ Universit\`{a} del Piemonte Orientale, Vercelli, Italy\\
$^{130}$ Universit\`{a} di Brescia, Brescia, Italy\\
$^{131}$ Variable Energy Cyclotron Centre, Homi Bhabha National Institute, Kolkata, India\\
$^{132}$ Warsaw University of Technology, Warsaw, Poland\\
$^{133}$ Wayne State University, Detroit, Michigan, United States\\
$^{134}$ Westf\"{a}lische Wilhelms-Universit\"{a}t M\"{u}nster, Institut f\"{u}r Kernphysik, M\"{u}nster, Germany\\
$^{135}$ Wigner Research Centre for Physics, Budapest, Hungary\\
$^{136}$ Yale University, New Haven, Connecticut, United States\\
$^{137}$ Yonsei University, Seoul, Republic of Korea\\
$^{138}$  Zentrum  f\"{u}r Technologie und Transfer (ZTT), Worms, Germany\\
$^{139}$ Affiliated with an institute covered by a cooperation agreement with CERN\\
$^{140}$ Affiliated with an international laboratory covered by a cooperation agreement with CERN.\\

\end{flushleft} 
\end{document}